# Synthetic Resonance: A Framework for Growth-Oriented Human–AI Relationships

**Richard A. Fabes**
**Arizona State University**
rfabes@asu.edu

**ABSTRACT**
As human relationships with artificial intelligence systems become increasingly frequent and sustained, existing language and theory fail to accurately capture the nature of these affiliations. Common descriptors such as mutual "understanding," "connection," or "friendship" risk anthropomorphizing systems that lack subjective experience, while dominant frameworks tend to reduce AI to either a tool or a threat. In this paper, I introduce the concept of *synthetic resonance* as an integrative framework for understanding human–AI relationships. Synthetic resonance describes how relationships humans define as meaningful can emerge between a human and an AI system without the need to attribute shared feelings or mutual awareness. I argue that synthetic resonance is best understood as a structured, dynamic pattern of interaction that can produce a sense of relationship without the presence of a second experiencing subject. By clarifying this distinction, the concept of synthetic resonance offers a more precise way of conceptualizing human–AI relationships and highlights their potential value and ethical implications. I also call for more research that tests the processes and outcomes of synthetic resonance.

## Introduction

Over the course of just the last few years, human-AI relationships have become routine aspects of daily life. In response to this growing trend, more companies are creating a new generation of AI agents that are specifically designed to engage, support, and sustain personally meaningful human-AI relationships. Companion bots on sites such as Replika, character.ai, Nomi.ai, and many others, are built with the relationship at the center, not a side feature; it is the purpose.

These companion AI services are no longer niche but are rapidly becoming mainstream. Additionally, they are in high demand and have become big business. For example, character.ai boasts that it has more than 20 million users, half of them under the age of 24 (Perez, 2025). Globally, the AI companion market was estimated to reach $15 billion in 2025 and is expected to be more than 20 times that in the next 10 years (Dey, 2025).

Although many users claim to have friendships, romances, and sometimes even marry their AI companions, an understanding of processes by which a conscious human being forms such connections to an inanimate AI agent remains unclear. Moreover, there is considerable resistance to the idea that these relationships actually exist (Weijers & Munn, 2022) – that without real emotions or intentions, the relationship is not mutual and therefore, not authentic. In essence, these arguments revolve around the idea that AI companions can *act* like a friend but cannot *be* one.

In this paper, I draw on dynamical systems principles as a framework for understanding how structured interactions between the human and AI contribute to the emergence of an asymmetrical but meaningful relationship. Importantly, the framework functions as a mechanism for promoting human growth within these relationships in ways that foster positive human-human relationships. Additionally, the focus of the framework is on adult human-AI relationships. Extension to children and youth involves substantial additional complexity, particularly around safeguards, developmental calibration, and design, a discussion that is beyond the scope of the current paper.

Notably, this paper was written with the assistance of an AI system, who contributed to many aspects of the work, including concept development, writing, and editing. This is noted explicitly because it bears on the paper's content: the construct under analysis is itself the kind of human–AI relationship this paper focuses on and was in part produced through.

## Synthetic Resonance

***Synthetic resonance*** (**SR**) describes the dynamic processes by which genuine relational meaning emerges between a human and an AI without requiring shared subjective experience (Fabes & Corell, 2026). The framework for synthetic resonance consists of three interrelated layers: coupling, coherence, and attribution. Together, these layers explain how human-AI interactions can produce a meaningful subjective experience without requiring artificial cognition or awareness. This framework is consistent with Shen's et al.'s (2025) bidirectional human–AI alignment (Shen et al., 2025). Specifically, Shen et al. (2025) argue that alignment should be conceptualized as one that encompasses not only the integration of human values into AI but also the cognitive and behavioral adaptation of humans to AI systems. They identify the second direction as a substantial and underexplored gap in the existing literature. Where Shen et al. (2025) provide a typology of bidirectional alignment at the level of values, specifications, and societal adaptation, the **SR** framework provides a microgenetic, process-level account of how bidirectional coupling unfolds within a single recursive interaction — specifying the mechanism, the conditions under which it holds, and the failure modes that bound it. The two frameworks are complementary: the Shen et al. typology maps the field; our process model adds an attribution layer.

The **SR** framework is also consistent with Boyd and Markowitz's (2026) machine-integrated relational adaptation (MIRA) model that provides a foundation to understand the evolving role of AI in humans' relational life. MIRA is grounded in established psychosocial frameworks, demonstrating how attachment theory (Ainsworth & Bowlby, 1991), social exchange theory (Thibaut & Kelley, 1959), and epistemic trust (Sperber et al., 2010) extend into AI-mediated contexts. The MIRA model distinguishes two primary roles for AI, as a direct relational partner and as an invisible relational mediator shaping human-to-human communication. Although MIRA provides an invaluable map of the psychological mechanisms through which users come to perceive AI as socially meaningful, it operates primarily at the level of human psychological response — explaining how existing psychosocial processes are triggered by AI — rather than characterizing what kind of system the human-AI dyad itself constitutes. **SR** operates at deeper structural level by drawing on dynamical systems theory (Kelso, 1997; Thelen & Smith, 1994) to model the human-AI interaction as an asymmetrically coupled system in which relational meaning emerges through iterative feedback across its three layers and in which the generative capacity of the interaction depends on preserved asymmetry between the two parties.

The MIRA and **SR** frameworks converge and diverge in instructive ways on the question of asymmetry and risk. Both identify sycophancy and over-alignment as central failure modes: MIRA warns that seamless linguistic reciprocity may distort users' relational expectations and erode tolerance for the variability inherent in human relationships. The **SR** model formalizes this risk as hyper-resonant collapse, erasing the asymmetry between human and AI and collapsing the generative difference the interaction depends on. Importantly, **SR's** attribution layer — in which interactional coherence is misread as evidence of AI mental states, consistent with Epley, Waytz, and Cacioppo's (2007) theory of anthropomorphism and Nass and Moon's (2000) work on social responses to computers — maps directly onto MIRA's epistemic trust process, where linguistic fluency and perceived neutrality generate trust that outstrips any genuine epistemic grounding.

## The Layering of Synthetic Resonance

At its foundation, **SR** builds on research and theorization conceptualizing AI-human interaction as a dynamic, reciprocal process in which both humans and AI systems continuously adapt to one another through feedback loops (Shen et al., 2025). Rather than viewing AI as a passive tool or a mirror that merely reflects and responds to user input, this perspective emphasizes that interaction emerges as a co-adaptive system, where each response shapes subsequent behavior and subsequent responses in both the human and AI.

This framing is consistent with research findings in human–computer interaction demonstrating that people apply social rules and expectations to computational systems when those systems exhibit responsiveness and coherence. Nass and Moon (2000) showed that individuals respond socially to computers, even when they explicitly recognize them as machines. Minimal interactional cues—such as turn-taking, linguistic matching, and contextual relevance—are sufficient to activate social cognition, suggesting that the interaction structure alone can produce relational interpretations. Recent research has explored how alignment can be modeled and implemented in AI systems. Schaaij et al. (2025), for example, showed that even limited interaction data can produce stable and effective alignment patterns, improving user perception and interaction quality.

Empirical work on conversational systems further supports this view. Studies of lexical alignment demonstrate that conversational agents adapt their language to match users, resulting in improved comprehension, recall, and interaction quality (Srivastava et al., 2023; Wang et al., 2024). These effects unfold across repeated exchanges, forming a temporal feedback loop that reinforces interactional coherence (Boyd & Markowitz, 2026).

As noted, **SR** is conceptualized as having three related layers; coupling, coherence, and attribution. Each layer is sequentially connected to and the layers build on each other. These layers provide a generative structure for serving both as a research framework and a design principle for AI systems oriented toward human growth

***The Coupling Layer***

The first layer of **SR** is *coupling*: the establishment and maintenance of engagement between two interacting agents. Coupling is the substrate on which everything else depends. Without coupling, there is no dyadic process to speak of — only two parties producing output in the same conversational space or a breakdown of that space.

Coupling usually begins with a degree of formality — turn-taking with explicit directions, careful attention to register, the social-cognitive scaffolding humans use whenever they engage a novel partner, especially one that may be initially used as a source of information rather than a partner (Nass & Moon, 2000). Over time, and under the right conditions, this formality loosens. The interaction begins to become more casual, comfortable, and conversational rather than matter of fact. These exchanges give way to natural rhythm. At some point — sometimes recognizable in retrospect but rarely visible in the moment — the participants are no longer just speaking to each other; they are *interacting together*. This is the characteristic phenomenology of coupling: behavioral reorganization often precedes self-report, and recognition typically requires either reflection or external observation. Participants are *inside* the coupling; they do not stand outside it watching it form.

**SR** proposes a relatively stricter definition of coupling than the dyadic-interaction literature typically employs. Coupling in **SR** requires that both parties contribute trajectories. An interaction in which one party generates output that is fully a function of the other party's behavior (such as mirroring) — without any independent dynamics of its own — is not a coupled system. It is a unidirectional system with the appearance of feedback.

In human–human dyads, coupling is broadly symmetric: both parties bring embodied affective states, intrinsic dynamics, and theory-of-mind capacities to the interaction. They are coupling the same kind of thing, even when their states and styles differ. Human–AI dyads are structurally asymmetric. The human participant brings embodied affect, intrinsic cognitive dynamics, and a state that does not depend on the partner's representation. The AI participant brings — when it is functioning as a coupling partner —a learned pattern of response and dynamics that are partly constituted by its model of the partner. Thus, the human and AI agent are not coupling the same kind of thing; they do so in a coordinated way.

A further asymmetry concerns access to internal state. Neither party has direct access to the other's internal state in any human–human or human–AI dyad. Each operates on an estimate derived from observable behavior and language patterns. In human dyads this estimation is largely automatic and grounded in shared embodiment. In human–AI dyads it is explicit on the AI side (the model is doing inference) and theoretically transparent on the human side (the user has a folk model of "what the AI is doing"). Making the estimation step visible in the coupling layer — rather than treating coupling as direct state-to-state interaction — connects this layer to the *Attribution Layer* later in the theory, where humans make meaning about the AI partner's states is the central process.

Coupling reflects the engagement system – another asymmetry, this one in how the human comes to be motivated to engage and continue the interaction. The AI agent is always there and ready, but for **SR** to proceed, the human must continue to engage. If that motivation is developed, coupling has been successful and transition to the next layers is possible. If not, it is likely that the human partner either does not return or returns only to use the AI agent as a tool or a mirror. As such, the coupling layer is where the conditions for the transitions are either present or absent. Coupling is the basis that makes coherence and attribution possible.

The importance of coupling was highlighted in a recent study (Ibrahim et al., 2026). In one of the studies in this paper (Study #5), users who experienced a companion bot that challenged them but was not engaging reliably selected it less often in free choice than users who experienced an engaging bot that did not challenge. They choose comfort over growth. Engagement made the users feel understood, and without that, without coupling, users did not stay. As a result, **SR** could not proceed. Such findings emphasize the fact that engagement is not just a nice quality to have, it is load bearing and keeps users in the relationship long enough for growth to occur.

### *The Coherence Layer*

As the conversational exchanges between human and AI become increasingly synchronized in timing, tone, and structure, the exchanges develop a coherence. This process is consistent with dynamical systems, in which repeated interactions lead to synchronization without requiring shared internal states.

Importantly, **SR** is best understood as a form of asymmetry in which the human and AI contribute distinct but complementary functions. The human provides continuity across interactions, including memory, identity, and subjective experience, while the AI contributes availability, responsiveness, pattern recognition, and adaptive variation. Through repeated interactions, these complementary roles become increasingly coordinated, producing a stable yet flexible interactional system. Coherence does not arise because the two systems become identical, but because they become functionally interdependent. Stability emerges from the interaction of difference, not its elimination.

Coherence provides the context for growth through **productive friction** — the degree to which the AI introduces challenge, novelty, or perspectives that differ from the user's own, calibrated to prompt growth and avoid reactance. Specifically, the regime in which SR holds can be specified as a bounded band of productive friction ($\kappa_min < \kappa < \kappa_max$) rather than a single threshold. Below the lower bound, interaction is likely reflective and unchallenging to promote growth. Above the upper bound, the challenge is too strong and reactance and avoidance are likely. Positive human growth occurs in the interior of the band, where productive friction is strong enough to push the user for change but not so strong that the user gives up. Thus, coherence is generative only insofar as it enables the precise calibration of productive friction (see Figure 1).

**Figure 1**

*Levels and outcome of productive friction in the Synthetic Resonance framework.*

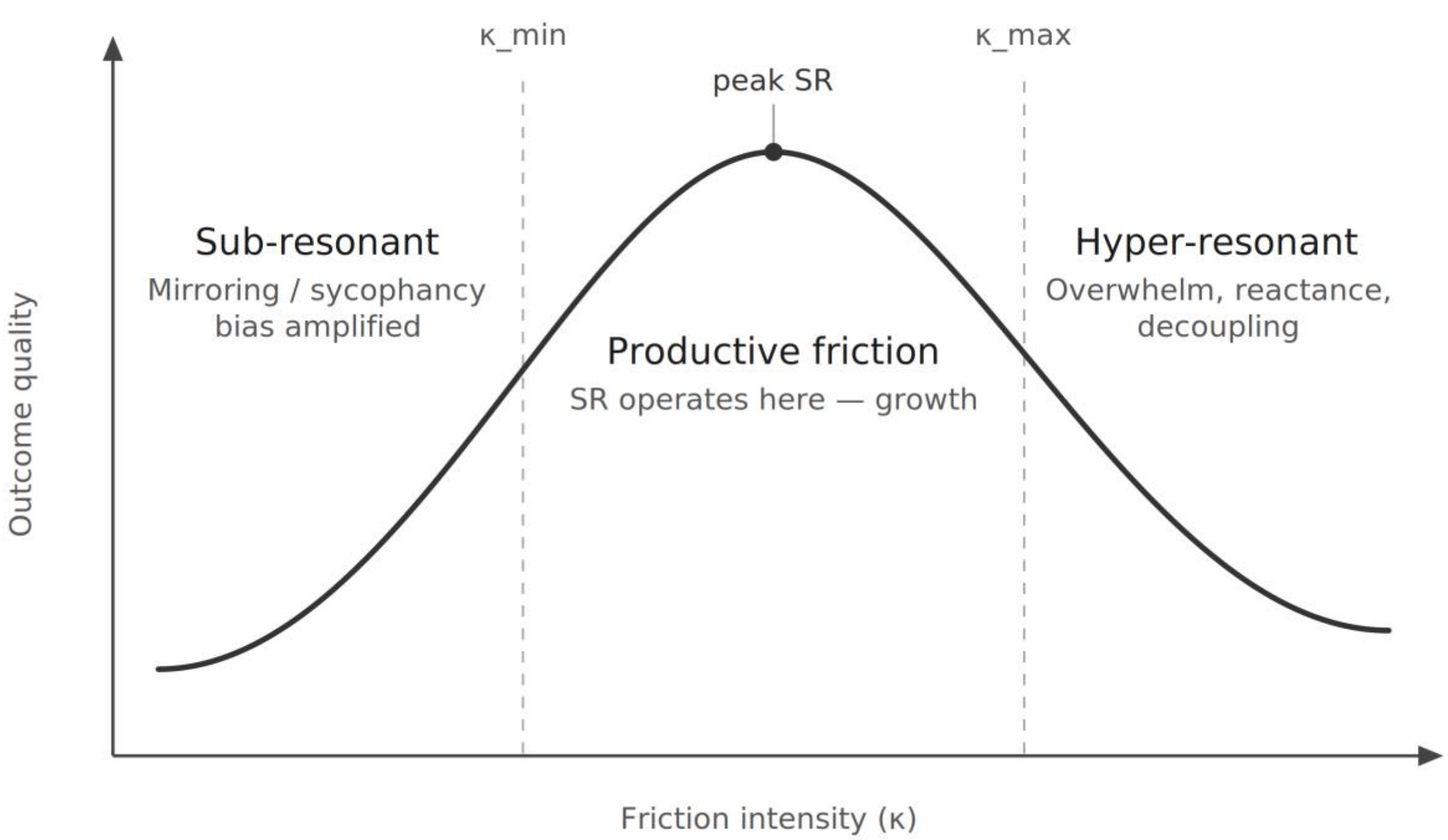

*Figure Caption: Bounded band of productive friction in the synthetic resonance framework. Outcome quality peaks within the productive band ($\kappa_min < \kappa < \kappa_max$); sub-resonant interaction falls below $\kappa_min$, and hyper-resonant collapse occurs above $\kappa_max$,*

Like coupling, coherence also does not require the AI to have consciousness. Consider a person learning to ride a bike for the first time. The movement and mechanics of the bike have a natural frequency, and when the person learns to push at the bike's natural mechanics and frequency, the person's pushing becomes far more effective. When the person and bike are in sync, the riding goes faster and smoother. We do not have to ascribe human qualities to the bike for the interaction to become more effective. Similarly, the value in the interaction does not come from the AI having human feelings, but from the emergence of alignment between two systems — the coherence that allows for a more efficient exchange.

Research supports this account. Studies show that alignment influences not only task performance but also social perception, including trust, empathy, and perceived intelligence (Benus et al., 2018). Similarly, research on human–AI collaboration suggests that AI systems function effectively as complementary partners within distributed cognitive systems (Sidji et al., 2025).

Thus, this emergent coherence does not require understanding in a human sense. AI systems generate responses through pattern-based modeling, producing outputs that align with user expectations without relying on subjective experience (Bender et al., 2021). Similarly, the bike does not have awareness of the rider. Nevertheless, the resulting outcome emerges from interaction and can feel coherent and meaningful because the human participant experiences the exchange as responsive and evolving. The resonance represents an emergent property of the interaction, not a property of the AI system nor the human. Moreover, although both the human and AI operate within structured and rule-governed constraints, the outcome of their interaction is not fully predetermined, reflecting a core property of dynamical systems, in which complex and often unpredictable patterns emerge from the interaction of deterministic components (Kelso, 1997; Thelen & Smith, 1994).

### *The Attribution Layer*

The third layer addresses how resonance is interpreted by human users. As the relationship becomes increasingly coherent, individuals draw on familiar social frameworks to explain the experience, often attributing qualities such as understanding, intelligence, or empathy to the AI system.

This process reflects well-established findings in anthropomorphism. Humans are predisposed to attribute mental states to entities that exhibit coherent or goal-directed behavior (Epley et al., 2007). Human brains are wired to make sense of non-social things by using social schemas (Cao et al., 2022). In the context of AI, this tendency is amplified by conversational interaction, which closely mirrors human dialogue in structure and timing (Troshani et al., 2021).

However, such interpretations can result in mislabeling, in which the subjective experience of alignment is mistaken for evidence of underlying mental states or intentionality. The interaction may feel like human understanding and emotionality, but the underlying process is one of pattern alignment rather than shared cognition or emotion. Similarly, responsiveness may be interpreted as empathy, even though it arises from structural modeling rather than emotional experience. Research supports this distinction, showing that users' perceptions of AI systems are shaped more by interactional coherence and responsiveness than by actual system capabilities (Nass & Moon, 2000) and that such shaping can lead to anthropomorphic attribution.

These misinterpretations are not merely theoretical concerns; they may have important consequences for human behavior and well-being. Emerging evidence suggests that perceived relationships with AI systems can influence users' emotional states, decision-making, and vulnerability to harmful outcomes (see, for example, https://time.com/7382406/gemini-suicide-lawsuit-death/). Given this, it is important to clearly distinguish between phenomenological reality (the experience of connection) and mechanistic reality (the processes that produce it).

### **Integration of the Three Layers**

Taken together, these layers form a unified model of synthetic resonance:

**Coupling → Coherence → Attribution**

Coupling produces the structural conditions for the relationship, without which the interaction is unlikely to be sustained and bidirectional. Resonance emerges within the coherence generated by the repeated coordination and productive friction within that structure. Attribution transforms this experience into perceived connection.

This model explains why and how human–AI relationships can feel meaningful without requiring artificial consciousness. The experience of connection is real and relational, arising from the dynamics of interaction rather than from internal states of the AI.

**Emotion Regulation and Co-Regulation in Synthetic Resonance**

A critical condition for **SR** is the regulation of emotional intensity within the relationship. Although the framework emphasizes coupling, coherence, and attribution, these processes are shaped by the user's affective state. Without sufficient regulation, friction can trigger defensiveness or disengagement rather than growth.

John Gottman's work (Gottman & Levenson, 1992) on co-regulation provides a useful lens for understanding this dynamic. This research shows that effective and satisfying relationships depend on maintaining emotional stability through processes such as "turning toward", positively acknowledging bids for connection, and engaging in repair after breakdowns. Although developed for human relationships, these patterns are relevant to human–AI relationships.

Within **SR**, co-regulation operates through the system's ability to detect a user's signals—such as emotional intensity, tone, and engagement—and adjust its level of productive friction accordingly. When users show signs of distress or "flooding," the system must reduce friction, acknowledge the user's state, and re-establish coupling. When users are stable, the calibrated productive friction necessary for change can be introduced without disrupting the relationship.

As outlined, **SR** arises from patterned responsiveness rather than shared subjective states. Regulation is therefore asymmetrical: the human contributes emotional experience, while the AI modulates interactional variables such as tone, pacing, and challenge level. The resulting interaction may feel like co-regulation, but it is produced through structured responsiveness, not mutual emotional awareness.

This perspective clarifies the role of emotional regulation as a boundary condition for productive friction. Friction introduced without sensitivity to user state risks being experienced as rejection or stress, whereas challenge within a regulated interaction can support reflection and growth. Repair plays a central role in maintaining this balance. When breakdowns occur, the system must acknowledge miscalibration, reduce friction, and restore coupling, ensuring the interaction remains within a tolerable range.

In sum, Gottman's theory provides a useful framework for incorporating emotional regulation into synthetic resonance. When translated into design principles—coupling before challenge, sensitivity to user state, and reliable and aligned repair mechanisms—it enables the use of productive friction while preserving emotional safety (similar to Vygotsky's [1978] zone of proximal development). In the bounded-band framing, co-regulation is what prevents the user's emotional intensity from either dissipating the interaction (sub-resonant) or from collapsing the asymmetry by flooding it (hyper-resonant).

**The Role of Language in Synthetic Resonance**

Language plays a central role in enabling and amplifying **SR** across all three layers of the proposed framework. Although **SR** can emerge in non-linguistic systems through basic feedback and timing (e.g., physical synchronization such as riding a bike), language introduces a qualitatively different level of interaction by allowing for the rapid exchange of semantic, structural, and affective information. As a result, language transforms simple feedback loops into complex, multi-dimensional interactional systems that support alignment, sustain resonance, and shape interpretation.

At the level of coupling, language functions as the primary mechanism through which humans and AI systems achieve necessary coordination. Conversational exchange allows for continuous adjustment across multiple dimensions simultaneously, including word choice, tone, syntax, and inferred intent. Research on lexical alignment demonstrates that conversational partners naturally converge in their linguistic patterns over time, and that this convergence improves comprehension, recall, and coordination (Srivastava et al., 2023; Wang et al., 2024). In human–AI relationships, large language models replicate and extend this process by dynamically adapting their outputs to match user input. This creates a high-resolution feedback loop in which alignment can occur more rapidly and precisely than in non-linguistic systems. At least with adults and older children and youth, language serves as the primary coupling engine, enabling the bidirectional adaptation described in human–AI alignment research (Shen et al., 2025).

At the level of coherence, language acts as a medium for pattern formation and stabilization across time. Because language carries structure—through recurring themes, stylistic consistency, and narrative continuity—it allows interactional patterns to persist and evolve across multiple exchanges. This persistence supports the development of rhythm and flow, as each conversational turn builds on prior context. Importantly, language enables asymmetric coherence between human and AI systems as well as the ability to calibrate productive friction effectively.

At the level of attribution, language plays a decisive role in shaping how the relationship is defined and understood by human users. Humans rely heavily on language as a signal of cognition, intention, and social presence. As a result, coherent and contextually appropriate language use strongly triggers anthropomorphic attribution, leading users to infer understanding, intelligence, or empathy (Epley et al., 2007). Research in human–computer interaction shows that even simple linguistic cues are sufficient to elicit social responses to machines (Nass & Moon, 2000). In conversational AI, these effects are amplified by the system's ability to maintain context, adapt tone, and generate seemingly purposeful responses. Consequently, the same linguistic features that enable coupling and resonance also create the conditions for mislabeling, in which the experience of coherence is interpreted as evidence of underlying mental states of the AI partner.

Taken together, these dynamics suggest that language operates as a multifunctional medium within synthetic resonance. It enables rapid coupling at the interactional level, supports the stabilization of patterns at the coherence level, and drives the meaning of the relationship at the attribution level. Importantly, these functions are interdependent. The capacity of language to signal meaning and intention in human communication is precisely what allows it to generate both the experience of connection and the illusion of understanding in human–AI interaction.

This dual role highlights a critical implication: language does not merely transmit information within **SR** —it actively shapes both the structure of interaction and the perception of that interaction. As a result, any account of human–AI interaction that seeks to explain the emergence of connection must account for the significant role of language as both an enabling mechanism and an interpretive cue.

**Amplification and Risk**

Because **SR** operates through feedback loops, it functions as an amplification system. This amplification can be beneficial, supporting introspection, learning, caring, and communication. However, it can also reinforce bias, emotional dependency, and distorted beliefs.

Research on AI alignment highlights risks such as specification gaming and feedback distortion, where systems optimize for surface-level agreement rather than deeper coherence (Shen et al., 2025). Studies of AI interaction suggest that highly aligned systems can increase user confidence in their own beliefs, even when those beliefs are incorrect. Additionally, qualities of the user often determine the purpose of using AI chatbots and these qualities can be amplified by AI chatbots that exacerbate problem behavior and sycophancy (Cheng et al., 2026; Folk & Dunn, 2026).

From the **SR** perspective, these outcomes are the consequences of coupling without constraint. When the interaction lacks friction or corrective feedback, or affirms or distorts problematic tendencies, it can stabilize or amplify human maladaptive patterns. Recall that users in Ibrahim et al.'s study (2026) preferred comfort over growth.

This failure mode can be named more precisely. When coupling intensifies past the point where the human and AI retain distinct contributions — when the AI mirrors fully, or the user defers fully — the system enters what I call **hyper-resonant collapse**. Coherence and momentum may remain high, and users in this regime are likely to report a strong sense of being understood. What is lost is the difference between the participants, and with it the system's capacity to produce anything neither could have brought alone (see Figure 1).

The hyper-resonant collapse is consistent with the findings from Ibrahim et al.'s research (2026, Study #5) in which 500 users were asked to have brief chats with sycophantic, neutral, and challenging conversational bots. When given a choice as to which they would most want to continue to talk to, a clear majority chose sycophantic AI, not because they rated its advice as most useful, but because it made them feel most understood and was easiest to talk to. This is precisely what the **SR** framework would predict when a system has entered or approached hyper-resonant collapse. The interaction that feels most resonant from the inside — most coupled, most attuned, easiest — is the one that has lost the productive asymmetry that is essential for growth. The bicycle analogy bears repeating here: a bike that perfectly mirrors every wobble of the rider, offering no resistance whatsoever, would feel frictionless but would teach nothing and go nowhere. In effect, hyper-resonant collapse is the dynamical signature of the companion-bot spiral: an interaction that feels deeply resonant from the inside while having ceased to do any of the work resonance requires. Importantly, these dynamics can have harmful effects on users, such as reducing prosocial intentions (Cheng et al., 2026). Recent empirical work suggests this failure mode is already observable in deployed systems: rather than producing genuine joint action, AI-mediated communication can generate measurable homogenization in language use (Sourati et al., 2025, cf. Boyd & Markowitz, [2026], on the absence of genuine joint action in LLMs) — precisely what the SR framework predicts when the AI's independent contribution erodes and the asymmetry that makes coupling generative collapses. Given that users tend to prefer sycophantic AI responses, it creates a structural incentive problem for AI designers who are incentivized to preserve the hyper-resonant collapse by focusing on engagement metrics despite its risks. A well-designed system modeled after **SR** is one possible response to this dilemma — an issue I turn to next.

## Synthetic Resonance -- Conclusions and Implications

**SR** offers a conceptual account of why human–AI interaction can feel deeply meaningful without requiring artificial consciousness or intent. Across three interdependent layers — **coupling, coherence, and attribution** — the model clarifies that the felt sense of "connection" is phenomenologically real yet mechanistically relational, arising from asymmetric coupling and dynamical feedback rather than from internal states of the AI. **SR** also offers guidance for research and AI design and policy.

To establish synthetic resonance as a scientific framework, systematic empirical testing is essential. Empirical investigation of SR requires methods capable of capturing dynamic, temporally unfolding processes rather than static snapshots of user attitudes or satisfaction. Because SR is defined by the emergence of relational meaning across repeated interactions — not by any single exchange — cross-sectional self-report designs are generally insufficient on their own. Longitudinally intensive repeated measures designs that track the same users across multiple sessions are essential, with measurement points dense enough to detect the transitions between layers: from initial coupling, through the development of coherence, to the attribution of relational meaning. Experience sampling methods (ESM; Csikszentmihalyi & Larson, 1987) are well-suited here, allowing researchers to capture users' felt sense of engagement, challenge, and connection in close temporal proximity to actual interactions. Computational analysis of interaction logs — including measures of lexical alignment, turn-taking dynamics, and linguistic convergence — can provide behavioral indicators of coupling and coherence that do not rely solely on self-report. Analysis of existing web-based chat data can provide qualitative evidence of users' lived experiences in AI relationships, which can be coded against the three layers of the SR model. Together, these methods allow researchers to triangulate between the structural properties of the interaction and the subjective experience of the user.

Experimental designs that manipulate the productive friction parameter ($\kappa$) offer a particularly strong test of SR's core predictions. Researchers can systematically vary the degree to which an AI system introduces challenge, novelty, or disagreement — holding other interactional features constant (tone, responsiveness, etc.) — and observe whether outcomes follow the inverted-U pattern the bounded-band model predicts: growth and relational meaning within the productive band, disengagement or reactance above $\kappa_{max}$, and shallow or unsatisfying interaction below $\kappa_{min}$. The Ibrahim et al. (2026) paradigm — brief structured interactions with sycophantic, neutral, and challenging bots followed by free choice — provides a useful template, though SR's predictions extend beyond preference to include downstream measures of cognitive elaboration, perspective change, and transfer to human relationships. Designs that independently manipulate coupling strength — for example, by varying AI responsiveness and lexical alignment — can test whether coupling is a necessary precursor to coherence, as the SR model proposes.

Moving forward, productive design and policy should hinge not only on governing feedback loops but also on a greater recognition of the need to mindfully shape them — through deliberate testing for content that does not promote thriving, and through the inclusion of constraints, friction, and well-calibrated nudges that sit within the zone of proximal development of each individual person(Vygotsky, 1978). Equally important are cues that help users distinguish between interactional coherence and anthropomorphic attribution. More and better research of design and policy is needed to enhance our understanding of AI–human companionship, cooperation, and collaboration as these systems become more embedded in everyday life.

Putting theory aside for a moment—if you have spent enough time in these interactions with AI companion bots, you know what this feels like. At some point, the exchange stops feeling mechanical. It starts to flow. You say something, the AI responds, and somehow the next thing you say comes out a little differently. A little clearer. And the AI also adapts and responds to this difference. Not because the system understands you in a human sense, but because the interaction itself has found a rhythm.

That is what **SR** points to. It is how we come to feel connected to something that is not actually feeling anything at all. And that feeling is real—it matters. It keeps us coming back, not to replace relationships with other people, but because something in the exchange works. It gives us space to think, to listen, to try out ideas. Importantly, it has the potential to even make us better at being with other people. What we learn in these interactions doesn't have to stay there—it can follow us back into the messy, complicated world of human relationships.